# Integrated Photodynamic Raman Theranostics for Cancer Diagnosis, Treatment, and Post-Treatment Molecular Monitoring

Conor C. Horgan[1,2,3], Mads S. Bergholt[1,2,3,†], Anika Nagelkerke[1,2,3,#], May Zaw Thin[4], Isaac J. Pence[1,2,3], Ulrike Kauscher[1,2,3], Tammy L. Kalber[4], Daniel J. Stuckey[4], Molly M. Stevens[1,2,3,*]

[1]Department of Materials, Imperial College London, London SW7 2AZ, UK.
[2]Department of Bioengineering, Imperial College London, London SW7 2AZ, UK.
[3]Institute of Biomedical Engineering, Imperial College London, London SW7 2AZ, UK.
[4]Centre for Advanced Biomedical Imaging, University College London, London WC1E 6DD, UK.
[†]Current address: Centre for Craniofacial and Regenerative Biology, King's College London, London SE1 9RT, UK.
[#]Current address: University of Groningen, Groningen Research Institute of Pharmacy, Pharmaceutical Analysis, P.O. Box 196, XB20, 9700 AD Groningen, The Netherlands.
[*]Corresponding author: m.stevens@imperial.ac.uk

## Abstract

Theranostics, the combination of diagnosis and therapy, has long held promise as a means to achieving personalised precision cancer treatments. However, despite its potential, theranostics has yet to realise significant clinical translation, largely due the complexity and overriding toxicity concerns of existing theranostic nanoparticle strategies. Here, we present an alternative nanoparticle-free theranostic approach based on simultaneous Raman spectroscopy and photodynamic therapy (PDT) in an integrated clinical platform for cancer theranostics. We detail the compatibility of Raman spectroscopy and PDT for cancer theranostics, whereby Raman spectroscopic diagnosis can be performed on PDT photosensitiser-positive cells and tissues without inadvertent photosensitiser activation/photobleaching or impaired diagnostic capacity. We further demonstrate that our theranostic platform enables *in vivo* tumour diagnosis, treatment, and post-treatment molecular monitoring in real-time. This system thus achieves effective theranostic performance, providing a promising new avenue towards the clinical realisation of theranostics.

## Introduction

A theranostic approach to cancer treatment, whereby therapeutic and diagnostic modalities are integrated into a single system, offers the potential for patient- or tumour-tailored therapies to improve clinical outcomes[1]. Through the combination of an appropriate diagnostic modality and a suitable therapy, clinicians could modify treatment protocols based on real-time diagnostic feedback, thereby tuning therapies to the patient or lesion under examination. Successful theranostic cancer management, however, requires the careful selection of compatible cancer diagnosis and treatment modalities and their effective combination into a single system for clinical use[2,3]. This, understandably, presents a formidable challenge.

To meet this challenge, existing theranostic strategies have primarily relied on the use of multifunctional (typically inorganic) nanoparticles for the conjugation, entrapment, or intrinsic demonstration of diagnostic and therapeutic agents[2,4–6]. Theranostic nanoparticle systems have been developed to enable MRI, CT, PET, and fluorescence imaging, in addition to chemotherapeutic compound delivery, providing a large library of potential diagnostic/therapeutic combinations[7,8]. Numerous studies have demonstrated theranostic constructs based on nanoparticle systems including gold nanoparticles[4,9], carbon nanotubes[10], quantum dots[11,12], and upconversion nanoparticles[13,14]. In these systems, optical modalities such as fluorescence imaging and Raman spectroscopy are regularly combined with phototherapies such as photodynamic therapy (PDT) and/or photothermal therapy (PTT) due to the inherent complementarity that exists between these modalities[7,15,16].

Despite exciting developments, whereby single nanoparticle systems perform as both diagnostic and therapeutic constructs, there remain significant concerns that have thus far stymied clinical translation efforts[3,17]. Chief amongst these are ongoing safety concerns, namely due to the non-biodegradability and subsequent bioaccumulation of many nanoparticle systems, as well as toxicities they may display[18–

[21]. Indeed, the complex synthesis required to introduce multifunctionality into many such nanoparticle systems not only increases production costs and regulatory hurdles, but makes pharmacokinetic and pharmacodynamic studies more difficult, due to complex interactions that may exist between multicomponent structures and any encapsulated materials[22]. Further compounding these issues is the inherent trade-off that often exists between diagnostic and therapeutic modalities in terms of desired dosages and clearance rates[23,24]. While optimal diagnostic performance often requires minimal contrast agent with rapid clearance, therapeutics typically necessitates maximal tolerated dosages for as long as possible to achieve high response[3,22].

Here, we demonstrate a theranostic approach to cancer treatment that avoids the need for complex inorganic nanoparticles altogether. We employ a custom-built, multimodal fibreoptic probe to combine Raman spectroscopy for cancer diagnosis with PDT for optical theranostics. Raman spectroscopy is an optical diagnostic modality capable of identifying subtle biochemical differences between tissues through the application laser light and subsequent collection of inelastically scattered light from tissue[25,26]. The resulting biochemical spectral fingerprint has enabled real-time *in vivo* diagnoses of gastrointestinal cancers[27–29], skin cancer[30], breast cancer[31], and cervical cancers[32,33] with sensitivities and specificities of between 85% and 95%. Importantly, this diagnostic performance is achieved without the need for contrast agents such as nanoparticle systems, with diagnostic readouts based solely on the underlying biochemistry of the tissues. Similarly, PDT is an optical therapeutic modality that combines light, oxygen, and a photosensitiser to provide spatially and temporally controlled tumour destruction[34,35]. Following local or systemic administration, photosensitiser activation by light of a specific wavelength produces a photochemical reaction that generates reactive oxygen species resulting in controlled tumour destruction directly and indirectly[36,37]. Crucially, PDT photosensitisers are small molecular compounds with existing clinical approvals, and have to date been applied for the treatment of skin cancers, oesophageal cancer, and head and neck cancer, among others[38–41].

Here we show, through the choice of an appropriate Raman spectroscopy excitation wavelength and the careful selection of PDT photosensitisers, it is possible to achieve effective cancer theranostics. Through extensive *in vitro* characterisation, we identify suitable Raman spectroscopy and PDT parameters towards clinical implementation of this approach. Finally, we demonstrate the feasibility of our theranostic approach using a custom-built, multimodal fibreoptic platform for the diagnosis, treatment, and post-treatment molecular monitoring of colorectal xenograft tumours in an *in vivo* mouse model. Together, our results highlight a potential strategy for the clinical translation of theranostic cancer management concepts.

## Results

### Integrated Fibreoptic Raman Spectroscopy and Photodynamic Therapy

We have developed a fibreoptic theranostic platform, capable of the simultaneous delivery of up to three laser light sources (Figure 1a). This theranostic platform encompasses a fibreoptic probe (2 mm probe tip diameter) consisting of a central optical fibre for Raman spectroscopic excitation. Surrounding the central optical fibre are seven optical fibres, one for PDT excitation, one for alternate PDT excitation (at a third wavelength), and five for the collection of Raman spectroscopic signal. Incorporated into the tip of this multimodal fibreoptic probe are distal optical components including shortpass filters, notch filters, and a focusing lens to maximise the efficient collection of Raman scattered light and the colocalised delivery of Raman and PDT laser light for contact or short distance applications. Figure 1b illustrates an example workflow using our developed theranostic platform for the diagnosis, treatment, and post-treatment molecular monitoring of a cancerous lesion. Here, when used in conjunction with a suitable PDT photosensitiser, our multimodal probe enables theranostic operation to realise the goal of patient- or lesion-specific diagnosis at the molecular level, treatment, and importantly, post-treatment monitoring of the tumour response to PDT via biomolecular fingerprinting.

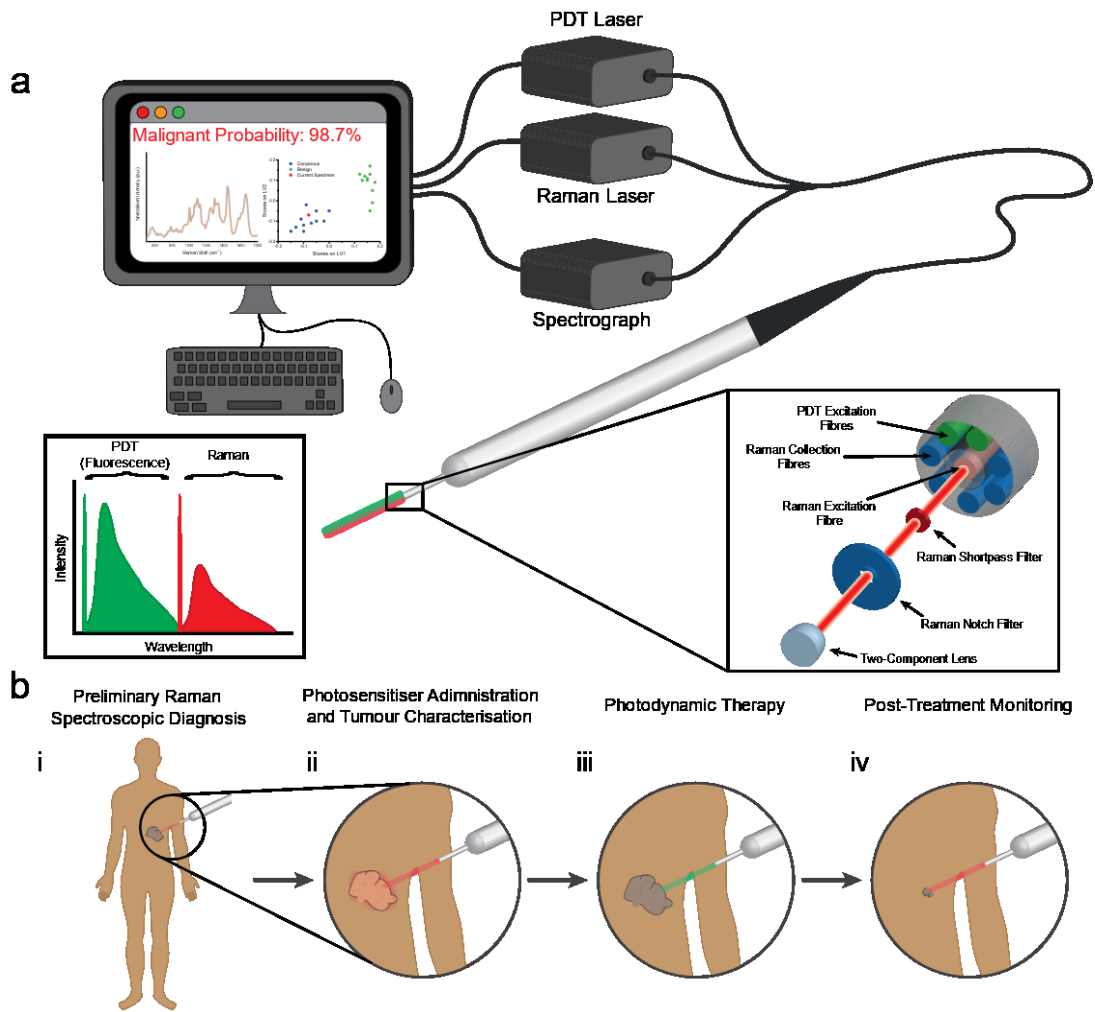

**Figure 1 | Multimodal Fibreoptic Probe for Nanoparticle-Free Optical Theranostics and Envisaged Surgical Workflow.** (**a**) Schematic of clinical system comprising a multimodal optical probe, a spectrograph, laser sources for Raman spectroscopy and photodynamic therapy, and a computer with associated system control software; (inset, left) Excitation-emission diagram demonstrating wavelength separation of diagnostic (Raman spectroscopy) and therapeutic (PDT) modalities. (**b**) Envisaged surgical workflow: (**i**), Raman spectroscopic identification of cancerous lesions; (**ii**) Photosensitiser administration resulting in preferential photosensitiser uptake in diseased tissues with no impact on Raman spectroscopic diagnostic capabilities; (**iii**) Activation of photosensitisers in target lesions through illumination with PDT laser; (**iv**) Post-treatment monitoring of treated areas demonstrating destruction of cancerous lesions.

**Photosensitiser Selection and Characterisation**

We first examined three clinically employed photosensitisers with excitation and emission profiles expected to be compatible with the 785 nm laser used for *in vivo* Raman spectroscopic diagnostics (Supplementary Figure 1). 5-aminolevulinic acid (5-ALA) is a prodrug to the photosensitiser protoporphyrin IX (PPIX), both of which occur naturally at low concentrations within almost all human cells, with clinical approval for PDT of actinic keratosis, basal cell carcinoma, Bowen's disease, and bladder cancer[42–44]. PPIX is excited at 633 nm for PDT, with fluorescence emission between 600 nm and 700 nm (Supplementary Figure 1a, d). The second, verteporfin is a clinically-approved photosensitiser with an excitation peak at 690 nm and fluorescence emission between 600 nm and 800 nm (Supplementary Figure 1b, e)[45,46]. Though originally developed for cancer applications, verteporfin is most commonly applied to the treatment of choroidal neovascularisation[47,48]. The third photosensitiser, temoporfin, is activated at 652 nm with fluorescence emission between 630 nm and 750 nm (Supplementary Figure 1c, f)[49,50]. Temoporfin is clinically approved in Europe for the treatment of squamous cell carcinomas of the head and neck[38].

Each of these photosensitisers was selected as their clinical excitation wavelengths are far from the 785

nm wavelength used for Raman spectroscopic excitation, and their fluorescence emissions fall well below the wavelength range from Raman spectroscopic signal collection. As expected, excitation of each of these photosensitisers at 785 nm produced almost no detectable fluorescence signal (less than 0.2% of the peak fluorescence signal upon photosensitiser excitation at 405 nm), indicating their potential compatibility with Raman spectroscopy (Supplementary Figure 1g-i). As a further initial screening of these photosensitisers, we investigated the background fluorescence they generate in Raman spectra when measured in solution using our multimodal fibreoptic probe (Supplementary Figure 2). In each case, at photosensitiser concentrations in solution of up to 1000 ng/mL (1.78 µM PPIX, 1.47 µM temoporfin, 1.39 µM verteporfin), no significant detectable increase in fluorescence background due to the photosensitiser was observed, indicating Raman spectroscopic diagnostics could likely be performed on tissues with comparably high photosensitiser concentrations. Indeed, work quantifying PPIX levels present in high-grade gliomas (which demonstrate very high PPIX accumulation) following 5-ALA application for fluorescence-guided surgery has previously indicated a mean concentration of 5.8 µM[51]. We anticipate that with the added scattering effects of tissues, Raman spectroscopy of lesions with similar PPIX concentrations would be possible.

**Compatibility of Raman Spectroscopy and Photodynamic Therapy *In Vitro***

The successful combination of Raman spectroscopy and PDT for cancer theranostics relies on a lack of interference between the two modalities. Firstly, it is essential to demonstrate that the laser used for Raman spectroscopy does not cause undesired premature activation of the photosensitisers employed for PDT. Inadvertent activation of photosensitisers could result in damage to healthy tissue and/or photobleaching of the photosensitiser, limiting the efficacy of PDT treatment on diseased tissues. Secondly, the intrinsic fluorescence of the photosensitisers must not impact or occlude the Raman spectral information obtained. Raman scattering is inherently weak and is therefore easily masked by stronger fluorescence signals[52]. In order to effectively perform molecular Raman diagnostics, a clear, strong signal is required for spectral discrimination of different pathologies.

To assess the compatibility of these two modalities, we first employed cell viability assays to investigate whether the 785 nm laser wavelength used for *in vivo* Raman spectroscopy activates any of the photosensitisers when tested at clinically relevant concentrations (Figure 2a-c and Supplementary Figure 3a-f). CCK-8 cell viability assays were performed on three different cell lines, A549 lung carcinoma cells, MDA-MB-231 breast adenocarcinoma cells, and MDA-MB-436 breast adenocarcinoma cells. In each case, illumination at the photosensitiser-specific wavelength resulted in photosensitiser dose-dependent cell death, while illumination at 785 nm did not affect cell viability relative to no illumination controls. These data thus indicate that Raman spectroscopy could be performed on photosensitiser-containing tissues *in vivo* without causing photosensitiser activation.

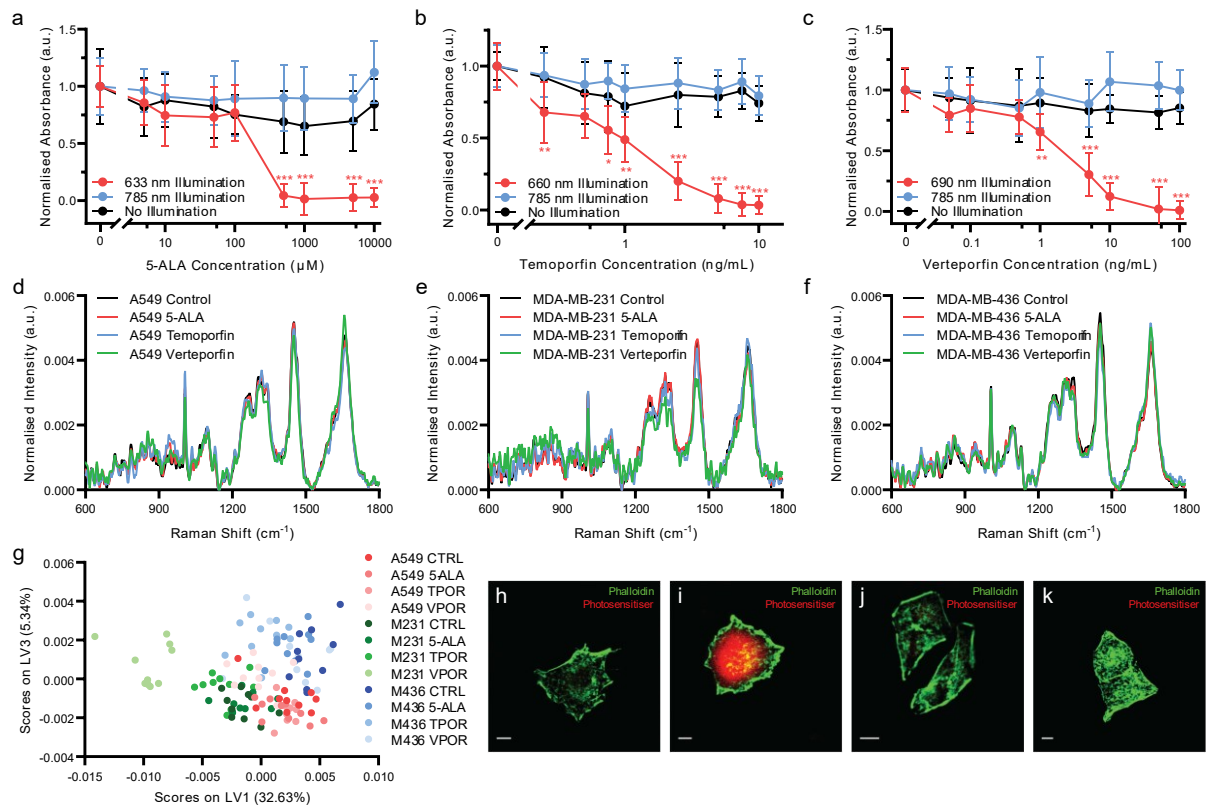

**Figure 2 | Compatibility of Raman Spectroscopy and Photodynamic Therapy *In Vitro*.** (**a-c**) CCK-8 cell viability assay of A549 cells incubated with (**a**) 5-ALA (PPIX prodrug), (**b**) Verteporfin, or (**c**) Temoporfin, at varying concentrations and illuminated with either the photosensitiser-specific LED array (633 nm, 690 nm, or 660 nm), the 785 nm LED array, or not illuminated (mean ± S.D., N = 3, n = 6) (Multiple comparisons *t*-test, Bonferroni post hoc correction, * $P < 0.05$, ** $P < 0.01$, *** $P < 0.001$). (**d-f**) Raman spectral acquisitions (10 s integration time) of cells in the presence of different photosensitisers (phenol red-free DMEM (Control), 5-ALA (10000 μM), Verteporfin (100 ng/mL), or Temoporfin (10 ng/mL)) (N = 10, n = 5). (**d**) A549 cells, (**e**) MDA-MB-231 cells, (**f**) MDA-MB-436 cells. (**g**) Partial Least Squares – Discriminant Analysis (PLS-DA) of Raman spectra from three different cell lines (A549, MDA-MB-231, and MDA-MB-436) incubated with one of the three photosensitisers or no photosensitiser, performed across cell line (N = 40, n = 5 for each cell line) with classification accuracies of A549: 98.3%, MDA-MB-231: 100.0%, and MDA-MB-436: 98.3%. (**h-k**) Confocal fluorescence images of photosensitiser positive A549 cells (**h**) Control, (**i**) (5-ALA induced) PPIX, (**j**) Verteporfin, (**k**) Temoporfin (scale bars 10 μm).

Next, to determine whether photosensitiser fluorescence impacts the Raman spectral information obtained, we collected Raman spectra using a confocal Raman microspectroscopy system at 785 nm from each of the three cell lines without photosensitiser and with each photosensitiser at the maximal dose tested for the cell viability assays (Figure 2d-f). For each cell type, the Raman spectra appeared grossly similar, with no substantial occlusion of the Raman spectral signal as would be expected for compounds with fluorescence emission in the Raman spectral range. Raman difference spectra between the photosensitiser positive cells and the control cells demonstrated only subtle spectral differences, potentially due to increased background noise induced by photosensitiser fluorescence (Supplementary Figure 4). Indeed, the raw Raman spectra for each cell line did show some changes in background fluorescence, with particularly notable increases in background fluorescence for MDA-MB-231 Temoporfin and Verteporfin cells (Supplementary Figure 5). However, in each case, the overall shape of the Raman spectrum and the key spectral peaks were maintained. This was further confirmed through assessment of the mean spectral coefficient of variation and the signal-to-noise ratio (SNR) (calculated as the peak intensity at 1650 cm$^{-1}$ divided by the mean standard deviation between 1780-1820 cm$^{-1}$) for each spectrum. These indicated generally consistent values across the processed Raman spectra, with increases in the mean spectral coefficient of variation and decreases in the SNR observed in the raw Raman spectra of the Temoporfin and Verteporfin cells (Supplementary Figure 6).

Partial least-squares – discriminant analysis (PLS-DA) provided further demonstration that the fluorescence of the photosensitisers investigated did not impact the Raman spectral information acquired (Figure 2g and Supplementary Figure 7). This analysis, performed using a Venetian blinds cross-validation, successfully classified cell spectra according to their respective cell phenotype with accuracies of greater than 98% (A549: 98.3%, MDA-MB-231: 100.0%, MDA-MB-436: 98.3%) irrespective of the presence of a photosensitiser within the cell. This provided further evidence that the processing performed on the raw Raman signal effectively accounted for minor spectral variations due to the presence of photosensitisers within the cells.

Lastly, the presence of photosensitisers in cells and their fluorescence emission upon blue light (405 nm) excitation was observed using confocal fluorescence microscopy (Figure 2h-k). Here, while fluorescence emission of PPIX was readily observable, verteporfin fluorescence was very weak while Temoporfin fluorescence was not detectable, despite photo-toxicity at this concentration.

Owing to the high fluorescence yield and low Raman spectroscopic fluorescence background of PPIX (Figure 2i and Supplementary Figure 5), which offers the potential for tumour imaging and treatment monitoring *in vivo*, we investigated the use of clinically-relevant Raman spectroscopy and PDT lasers through our multimodal fibreoptic probe with 5-ALA induced PPIX *in vitro*. LIVE/DEAD$^{TM}$ assays of cells incubated with 5-ALA indicated extensive, highly localised cell death at the point of 633 nm laser application whilst minimal cell death was observed for the 785 nm laser and no illumination control cells despite 785 exposure being 34 times greater than the maximum permissible skin exposure limit of 1.63 W.cm$^{-2}$ defined by the American National Standards Institute (Figure 3a-c)[53,54]. In the absence of 5-ALA induced PPIX, minimal cell death was observed for all three conditions (Figure 3d-f).

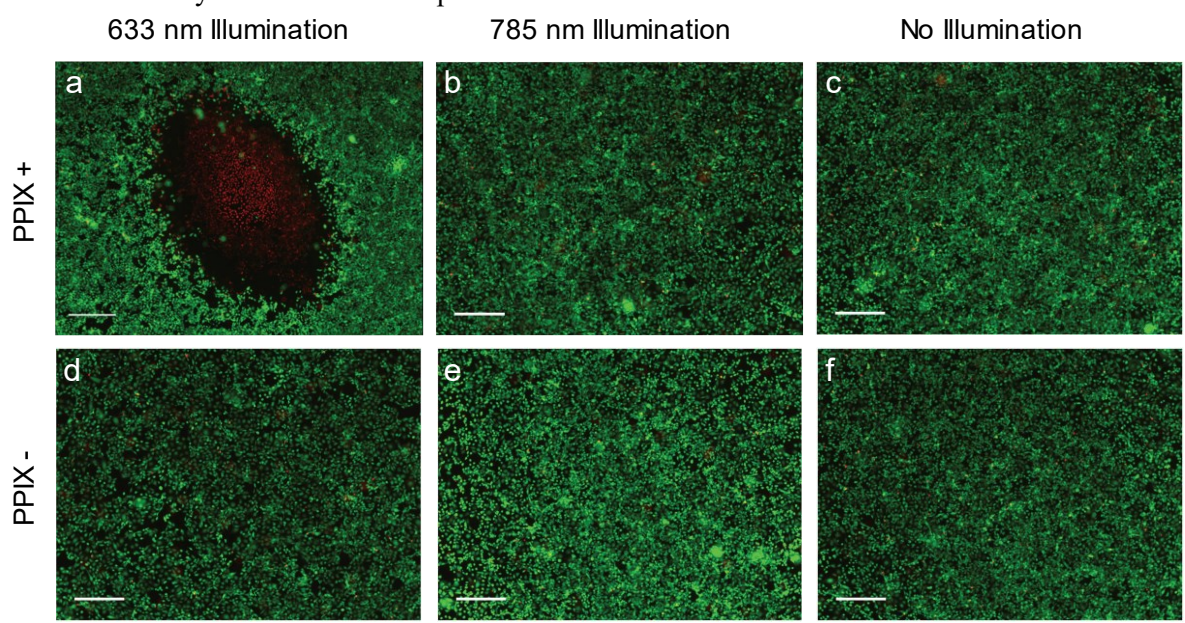

**Figure 3 | Compatibility of Raman Spectroscopy and Photodynamic Therapy Multimodal Fibreoptic Probe *In Vitro*.** (a-f) LIVE/DEAD$^{TM}$ stain of A549 cells incubated with 5-ALA/PPIX and exposed to laser light at 100 mW for 120 seconds (total exposure 34 W.cm$^{-2}$, maximum permissible skin exposure 1.5 W.cm$^{-2}$) (scale bars 400 μm). (**a**) 5-ALA/PPIX, 633 nm laser (multiple images manually stitched together), (**b**) 5-ALA/PPIX, 785 nm laser, (**c**) 5-ALA/PPIX, no illumination, (**d**) Control, 633 nm laser, (**e**) Control, 785 nm laser, (**f**) Control, no illumination.

Taken together, the *in vitro* results presented here indicate a high level of compatibility between 785 nm Raman spectroscopy and the three photosensitisers examined. Of these photosensitisers, 5-ALA induced PPIX demonstrated the lowest impact on Raman spectra, ready fluorescence observation under blue light excitation, and has the widest range of clinical PDT approvals, making it a clear choice to take forward for *in vivo* investigations.

### *In Vivo* Cancer Theranostics

To demonstrate the potential of our theranostic approach for cancer, we next conducted an *in vivo* study of SW1222 colorectal tumour xenografts in *nu/nu* mice. Ten female *nu/nu* mice with SW1222 tumours on their right flank following subcutaneous injection with $1\times10^6$ SW1222 cells, engineered to express the firefly luciferase gene for bioluminescence image (BLI), were split into a control group and a PDT group (N = 5 in each group). Raman spectra were acquired (785 nm, 100 mW, 1 s) from the tumours and control flanks of both PDT mice and control mice prior to and 4 h post 5-ALA administration in PDT mice (Figure 4). Prominent Raman peaks were observed for both the tumour and control tissue at 936 cm$^{-1}$ (ν(C–C) proteins), 1078 cm$^{-1}$ (ν(C–C) of lipids), 1265 cm$^{-1}$ (amide III ν(C–N) and δ(N–H) of proteins), 1302 cm$^{-1}$ (CH$_2$ twisting and wagging of lipids), 1445 cm$^{-1}$ (δ(CH$_2$) deformation of proteins and lipids), 1655 cm$^{-1}$ (amide I ν(C=O) of proteins), in line with previous *in vivo* Raman spectroscopic studies[27,28,55]. Significant Raman spectral differences between the control and SW1222 tumour tissue were observed both before and after 5-ALA administration, with greater peak intensities in the control tissue as compared to the tumour tissue at 1265 cm$^{-1}$, 1302 cm$^{-1}$, 1445 cm$^{-1}$, and 1655 cm$^{-1}$ (Figure 4c). Notably, we observed an up-regulation of protein content in tumour tissue, as indicated by an increase of the 1655 cm$^{-1}$ peak relative to the 1445 cm$^{-1}$ peak as well as a slight broadening of the 1655 cm$^{-1}$ peak relative to the control tissue, in line with previous Raman spectroscopic studies[27,55]. An increase in background fluorescence intensity was observed in the raw Raman spectra 4 h post 5-ALA administration for both control and SW1222 tumour tissue (Supplementary Figure 8a). However, upon spectral processing much of these differences were accounted for such that visual discrimination of spectra taken prior to and 4h post 5-ALA administration was difficult (Figure 4c). PLS-DA of processed Raman spectra from control and SW1222 tissues (performed across control and SW1222 tumour tissue, irrespective of PPIX presence) demonstrated a cross-validation accuracy of 97.4%, indicating the feasibility of highly accurate Raman diagnostics on PPIX-containing tissues (Figure 4d, Supplementary Figure 9), as indicated by previous studies[56]. Interestingly, despite being performed across tissues irrespective of PPIX presence, the PLS-DA successfully separated tumour tissue pre-PPIX and 4 h post 5-ALA, while failing to perform the same separation for the control tissues. Analysis of the mean spectra and PLS-DA latent variables indicate the influence of the peak at 1550 cm$^{-1}$ in the separation of pre- post-PPIX tumour tissue, with this peak notably having previously been attributed to porphyrins[57]. Confirmation of PPIX tumoural accumulation was performed for control SW1222 tumours and PDT SW1222 tumour tissues following re-administration of 5-ALA 6 days post PDT (Supplementary Figure 8b-c).

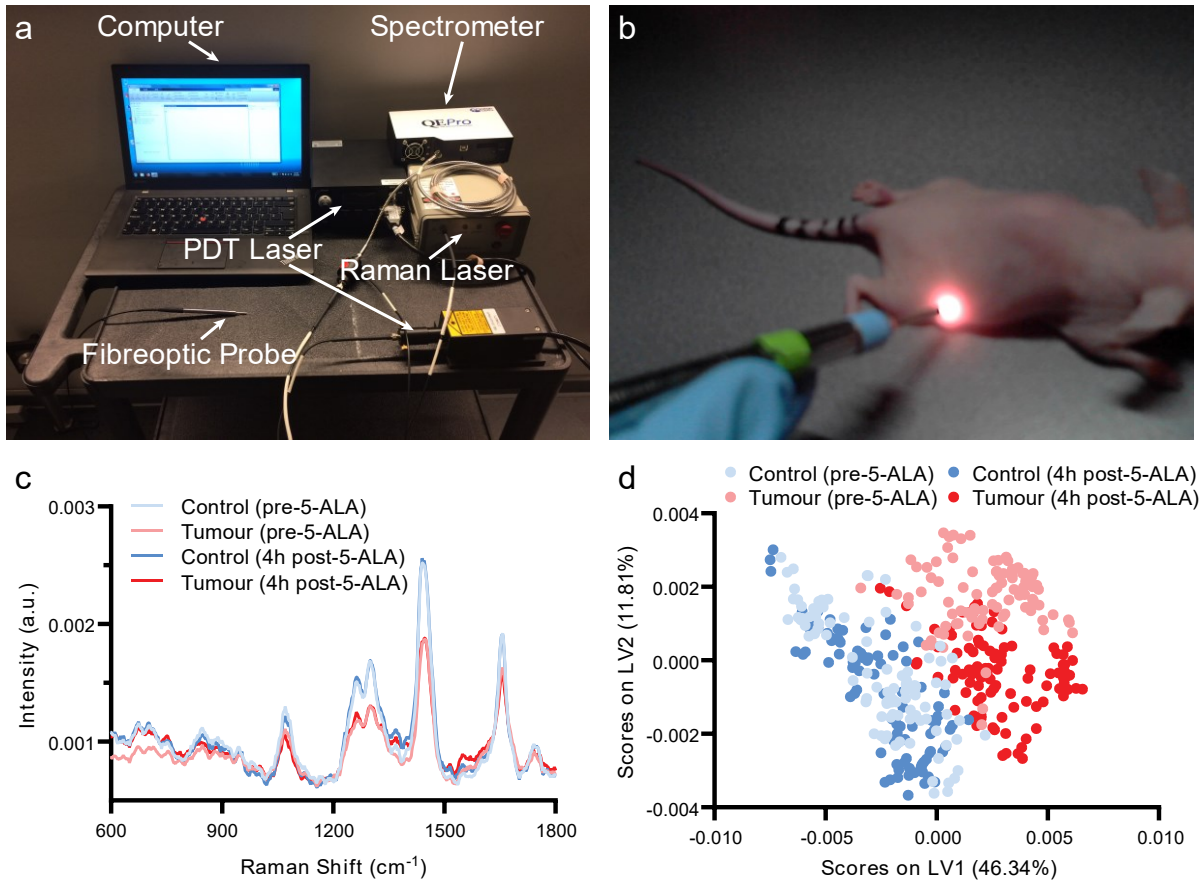

**Figure 4 | Raman Spectral Diagnostics of PPIX Positive SW1222 Tumours *In Vivo*.** (**a**) Integrated fibreoptic photodynamic Raman theranostic system. (**b**) In vivo Raman spectral acquisitions from nude mouse. (**c**) Mean processed Raman spectra of control flanks and tumours in mice pre-5-ALA induced PPIX and 4 hours post-5-ALA injection (50 mg/kg) (n = 18-20, N = 5). (**d**) Partial Least Squares – Discriminant Analysis of processed Raman spectra from control flanks and tumours performed across control and SW1222 tumour tissue irrespective of photosensitiser presence (n = 18-20, N = 5).

Lastly, we evaluated the remaining two steps of our proposed theranostic workflow (Figure 1b). At 4 h post 5-ALA administration, PDT mice were exposed to 633 nm laser light, applied through the same multimodal fibreoptic probe used for Raman spectroscopic diagnosis, with a fluence rate of 50 mW/cm$^2$ and a total fluence of 30 J/cm$^2$. Here, using a comparatively low PDT dose, we first examined whether PDT-mediated tumour control was possible following Raman spectroscopic diagnosis of PPIX-containing tissues and secondly, whether Raman spectroscopy could be used to perform post-treatment monitoring of the affected tissues. BLI imaging of control and PDT tumours prior to 5-ALA administration/PDT, at 3 days post-PDT, and at 6 days post-PDT demonstrated a significant reduction in BLI signal, indicative of reduced tumour growth at 6 days post-PDT relative to control tumours (one outlier PDT tumour excluded as determined by the 1.5x interquartile range (IQR) test, 22.2% normalised tumour growth at 3 days) (Figure 5a-b). Importantly, these data demonstrate that significant photosensitiser photobleaching did not occur following Raman spectroscopic diagnosis of PPIX-containing tissues. While there was no statistically significant difference in excised tumour mass for either group, 53.4 ± 37.6 mg for the control group vs. 52.8 ± 18.8 mg for the PDT group (mean ± STD, n = 5), BLI results did align well with post-treatment Raman spectroscopic molecular monitoring. Sequential Raman spectral analysis, performed via difference spectra between tumours and corresponding control tissue at each timepoint, initially indicated larger discrepancies for the PDT tumours than for the control tumours prior to PDT (day 0). However, at 3- and 6 days post-PDT the opposite was true, with PDT tumour discrepancies decreasing, while control tumour discrepancies increased across the 1265 cm$^{-1}$, 1302 cm$^{-1}$, 1445 cm$^{-1}$, and 1655 cm$^{-1}$ peaks (Figure 5c). We hypothesise that these spectral changes observed in the PDT tumours correspond to a small reduction in tumour load (i.e. the local density of tumour cells), resulting in a lower proportion of tumour cells in the Raman sampling volume. This thus suggests the feasibility of a Raman spectral method for post-treatment response monitoring at the molecular level and points to the potential of our single-device approach to diagnosis, treatment, and post-treatment monitoring.

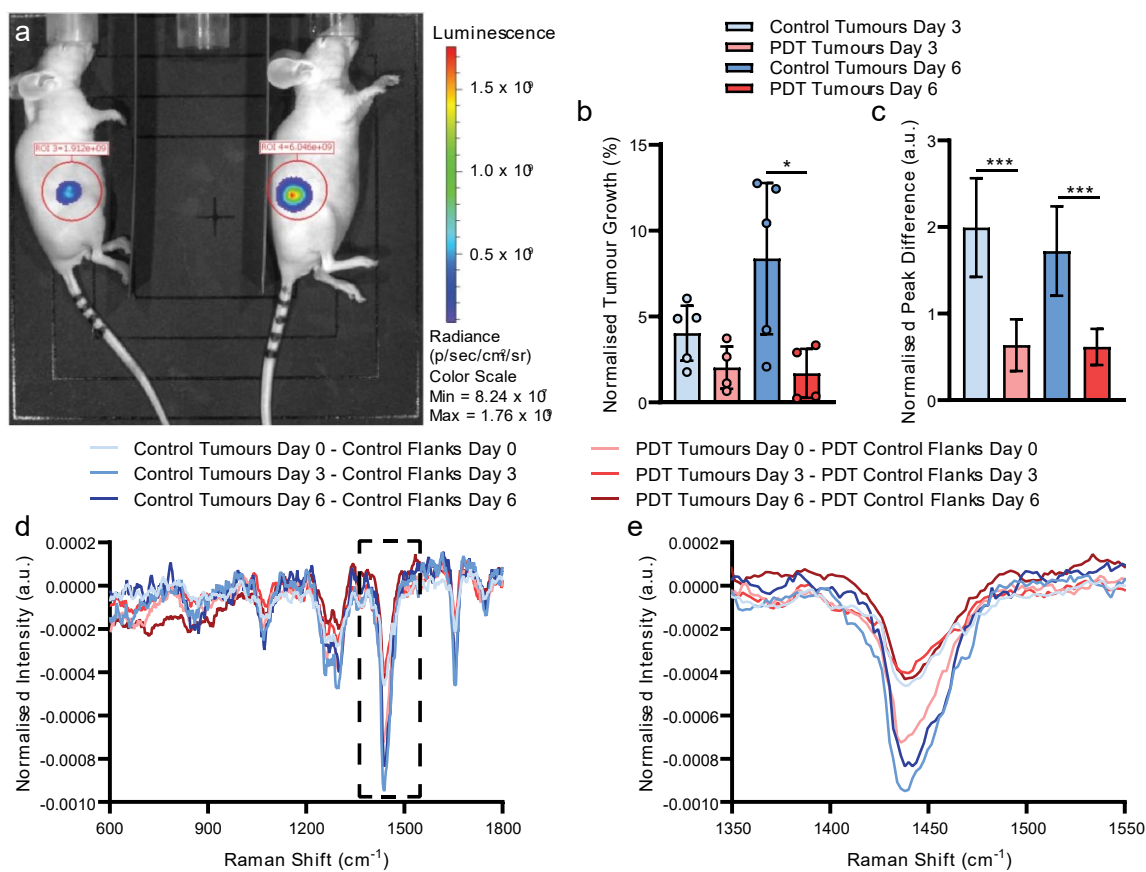

**Figure 5 | BLI and Raman Monitoring of PDT Efficacy for SW1222 Tumours.** (**a**) Exemplar BLI image used for assessment of tumour size. (**b**) BLI-measured tumour growth for control (N = 5) and PDT (N = 4, 1 outlier excluded as determined by the 1.5x IQR test, 22.2% normalised tumour growth at 3 days) mice at 3- and 6-days post-PDT (normalised to BLI-measured tumour sizes prior to PDT) (data are shown as mean +/- STD) (two-tailed Welch's t-test, * $P < 0.05$). (**c**) 1440 cm$^{-1}$ peak intensity difference for control and PDT mice calculated from the difference spectra shown in (**d** & **e**) at day 3 and day 6, normalised to day 0 (data are shown as mean +/- STD) (two-tailed Welch's t-test, *** $P < 0.005$). (**d**) Mean Raman difference spectra between tumour tissue and control tissue for control (n = 20, N = 2) and PDT (n = 20, N = 5) mice before (day 0) and 3- and 6-days post-PDT. (**e**) Magnified view of difference spectra shown in (**d**) between 1350-1550 cm$^{-1}$.

## Discussion

Theranostic approaches to cancer management offer the potential for patient- or tumour-tailored therapies based on real-time diagnostic feedback. Striving for this goal, theranostic systems have become increasingly advanced, incorporating an array of diagnostic and therapeutic modalities such as CT, PET, MRI, and photoacoustic imaging, as well as laser ablation, and photo- and chemotherapies and into various theranostic nanoparticle constructs[8,58–61]. However, despite these developments in theranostic platforms, clinical translation continues to be stymied by ongoing concerns over the safety and efficacy of the nanoparticle constructs on which they rely[18,62].

The goal of this work was thus to develop an optical theranostic system that avoided the need for nanoparticles altogether. To this end, we developed a theranostic approach through application of a fibreoptic probe. Owing to the recent development of small, portable fibreoptic probes[63–65], accurate, real-time Raman spectroscopic cancer diagnosis *in vivo* is readily achievable. By employing a custom-built fibreoptic probe, we combined Raman spectroscopic diagnosis with PDT for cancer theranostics. Crucially, this approach enables us to avoid many of the pitfalls of existing theranostic systems.

To investigate the clinical potential of this approach, we first detailed the complementarity of Raman spectroscopy and PDT. We showed that through careful material and parameter selections, Raman spectroscopic classification of photosensitiser-positive cells and tissues is feasible without inadvertent photosensitiser activation or photobleaching. Through *in vitro* experimentation, we identified PPIX as a particularly strong candidate for nanoparticle-free theranostics *in vivo*. Lastly, we demonstrated our optical theranostic approach using the custom-built, fibreoptic probe for colorectal xenograft tumours in an *in vivo* mouse model. We demonstrated that, using this approach, we could effectively perform accurate diagnosis, enable tumoural control, and achieve post-treatment response monitoring at the molecular level. Together these data thus highlighted the potential of our theranostic platform as an alternative strategy towards theranostic clinical translation.

Our method offers four key advantages over existing theranostic platforms. Firstly, it is unaffected by the significant toxicity and regulatory concerns levelled at inorganic nanoparticles[18,66]. Secondly, Raman spectroscopic diagnosis in this case is independent of the tumoural accumulation of photosensitisers (or other nanoparticles), removing concerns about nanoparticle targeting or imaging contrast for diagnosis (though photosensitiser tumoural accumulation remains essential for therapeutic purposes)[67]. Thirdly, the separation of diagnostic and therapeutic functions avoids trade-offs (e.g. circulation time, targeting, dosage) between modalities that hamper existing theranostic systems, limiting the efficacy of the individual components[3]. Finally, our system avoids the need for the complex, lengthy, and costly synthesis processes often required for the production of multifunctional theranostic nanoparticles, thereby reducing the associated costs and regulatory hurdles[22].

While our proposed theranostic system offers many advantages over existing theranostic platforms, it also faces several limitations. Firstly, as both Raman spectroscopy and PDT are light-based techniques, each is constrained by the penetration of light into tissue, making the system suitable only for surface lesions (i.e. early-stage cancers or pre-cancerous stages) or endoscopically-accessible lesions[17,67]. Secondly, while PDT has shown utility across many different cancers, treatment success is largely dependent on sufficient and preferential uptake of a photosensitiser into diseased tissue[17]. While development of increasingly powerful photosensitisers is ongoing, with a focus on increasing targeting efficacy and enabling combination therapies[68–70], treatment efficacy can be limited relative to other modalities. Lastly, although the safety and efficacy of Raman spectroscopic diagnosis has been successfully demonstrated across thousands of patients and many different cancers, clinical approval currently only exists for skin cancer applications, with work ongoing to realise Raman spectroscopic diagnosis for additional diseases[30,71].

In conclusion, we have developed a multimodal fibreoptic probe for nanoparticle-free cancer theranostics, combining Raman spectroscopic diagnostics with photodynamic therapy. We have successfully demonstrated the utility of this system for the *in vivo* diagnosis, treatment, and post-treatment monitoring of cancerous lesions. Importantly, our system achieves this without the need for complex, costly, and potentially toxic nanoparticles. This thus represents an alternative theranostic cancer management approach with improved clinical translation potential.

## Materials and Methods

### Fibreoptic Photodynamic Raman Theranostic Platform

The photodynamic Raman theranostic platform comprises five key components; a custom-built multimodal fibreoptic probe (EmVision LLC), a spectrograph (OceanOptics Inc), a laser source for Raman spectroscopy (785 nm, 600 mW, B&W Tek), a laser source for photodynamic therapy (633 nm, 150 mW, Roithner Lasertechnik), and a computer (Lenovo Thinkpad T460, Intel Core i5-6200U CPU) with custom-built software for system control. The probe tip uses a two-part lens to focus light from both the Raman spectroscopy and PDT lasers and is designed for close-proximity or direct contact Raman spectroscopic measurements (working distance: < 0.5 mm, 785 nm tissue penetration depth: ~ 1 cm in skin, but tissue-dependent)[72,73]. Raman spectroscopic diagnosis followed by PDT is thereby performed using a single probe, with lasers controlled through clinician-facing custom-built software.

### Photosensitiser Fluorescence Characterisation

Photosensitiser fluorescence emission was characterised using a Horiba FL-1000 spectrofluorometer. For each photosensitiser (Protoporphyrin IX (PPIX) (Sigma-Aldrich), Verteporfin (Sigma-Aldrich), Temoporfin (Biolitec)), two fluorescence spectra were acquired with 1 nm step and 5 nm slit widths. Emission spectra between 450 nm and 785 nm were acquired following excitation at 405 nm, while emis-sion spectra between 805 nm and 1000 nm were acquired following excitation at 785 nm.

10,000 ng/mL solutions of each photosensitiser were prepared in PBS from concentrated stock solutions. Five Raman spectra were collected from each solution using the custom-built multimodal fibreoptic Raman probe (EmVision) using a 785 nm Raman laser with 100 mW power output and 1 second integration time. Serial dilutions of each solution were thus prepared, and Raman spectra recorded with a Raman spectrum of PBS as a baseline.

**Cell Culture**

Cell experiments were performed using two human breast cancer cell lines (MDA-MB-231 and MDA-MB-436) (ATCC) and one human lung cancer cell line (A549) (ATCC). Cell lines were authenticated using STR profiling. Briefly, MDA-MB-231 and MDA-MB-436 cells were grown at 37 °C and 5% $CO_2$ in high glucose (4.5 g/L) DMEM GlutaMax (Life Technologies) supplemented with 10% (v/v) fetal bovine serum (FBS), 1x penicillin-streptomycin, 1x non-essential amino acids, and 20 mM pH 7.3 HEPES buffer solution. A549 cells were grown at 37 °C and 5% $CO_2$ in RPMI 1640 (Life Technologies) supplemented with 10% (v/v) FBS, 1x penicillin-streptomycin, 1x non-essential amino acids, and 20 mM pH 7.3 HEPES buffer solution. SW1222 human colon rectal cancer cells (ATCC) used for *in vivo* experiments were grown at 37 °C and 5% $CO_2$ in DMEM (Life Technologies), supplemented with 10% (v/v) FBS and 1x penicillin-streptomycin.

**Confocal Raman Spectroscopy**

$0.1 \times 10^6$ cells were seeded onto 25 mm diameter $MgF_2$ windows (Global Optics) in a 6-well plate and incubated for 24 hours. Cells were treated with the relevant photosensitiser in serum-free DMEM at the appropriate concentration and duration (1 hour, 100 ng/mL for Verteporfin; 3 hours, 10 mM for 5-ALA; 24 hours, 10 ng/mL for Temoporfin) and were then fixed with 4% paraformaldehyde (PFA) solution for 20 minutes at room temperature and stored at 4 °C in PBS until required for imaging. Cell spectra were obtained from cells in PBS using a 63x water-immersion objective lens and a 100 μm fibre acting as a confocal pinhole on a confocal Raman microscopy setup (Witec GmbH). Raman spectra were obtained using a 785 nm laser (Toptica Extra) with a power of ~80 mW and a 10 second integration time. For each cell, 5 spectra were obtained from random locations within the cell. Spectral processing was performed in MATLAB (2017b) using scripts developed in-house. First, the spectra were cropped to remove laser contribution and the fluorescence background was removed using a Whittaker filter baseline subtraction ($\lambda = 100,000$). Cosmic ray peaks present in the spectra were removed and the spectra were smoothed using a 1$^{st}$ order Savitzky-Golay filter with a frame width of 7. Partial least squares-discriminant analysis (PLS-DA) was performed using PLS Toolbox (Eigenvector Research) within the MATLAB environment. Pre-processed, normalised (area under the curve), and mean-centred spectra were classified using PLS-DA performed with 6 latent variables and a Venetian blinds cross-validation with 10 data splits.

**Confocal Fluorescence Imaging**

$0.3 \times 10^6$ A549 cells were seeded onto sterile glass coverslips in a 6-well plate in supplemented RPMI. 24 hours post seeding, RPMI was removed from the cells and cells were washed with DPBS. Cells were then incubated at 37 °C and 5% $CO_2$ in one of the following: serum-free DMEM for 1 hour; 100 ng/mL Verteporfin in serum-free DMEM for 1 hour; 10 mM 5-ALA in serum-free DMEM for 3 hours; 10 ng/mL Temoporfin in serum-free DMEM for 24 hours. Following incubation, photosensitiser solutions were removed and cells washed with DPBS. Cells were then fixed with 4% (v/v) PFA solution for 20 minutes at room temperature. After fixation, PFA solution was removed and cells washed twice with DPBS. Cell staining (under dark conditions) was performed as follows. DPBS was removed from cells and cells were incubated with 0.2% Triton-X in DPBS for 10 minutes at room temperature. Triton-X was removed from cells and cells washed three times with DPBS. Cells were then incubated with 5% (w/v) bovine serum albumin (BSA) (Sigma-Aldrich) for 1 hour at room temperature. BSA was removed from cells and cells washed twice with DPBS. Cells were then incubated with AlexaFluor 488-conjugated Phalloidin (Thermo Fisher), 1:40 in DPBS for 45 minutes at room temperature. Solution was removed and cells were washed 5 times for 5 minutes each time with 0.2% (v/v) Tween 20 (Sigma-Aldrich) in DPBS. Cells were then washed twice with DPBS. Coverslips were then removed from 6-well plate and mounted onto microscope slides (VWR) using Fluoromount (Serva) and allowed to dry before imaging. Imaging was performed using a Leica SP5 inverted confocal microscope equipped with a 405 nm diode laser and a 20x objective. Excitation was performed at 405 nm, with collection from 630 nm to 800 nm.

## LED Array Construction and Characterisation

Four single-colour LED arrays, consisting of 96 LEDs in a 12 x 8 parallel-series arrangement, were constructed using LEDs with wavelengths corresponding to photosensitiser activation or Raman imaging (631 nm - LED631E (ThorLabs), 660 nm - SSL-LX5093XRC/4 (Lumex), 690 nm - LED690-03AU (Roithner LaserTechnik GmbH), 780 nm - LED780 (ThorLabs). LEDs were arranged on a circuit board (RS Components) such that each LED illuminated a single well of a 96-well plate. LEDs were soldered in place with a resistor for each row of 8 LEDs to regulate the LED current loading. Optical power output was characterised using a PM100D optical power metre (ThorLabs), tuned to 633 nm, 660 nm, 690 nm, or 785 nm with the LEDs tuned to approximately 5 mW/cm$^2$ (3.5 mW/cm$^2$ for 660 nm LEDs).

## Cell Viability Assay

Using clear 96-well plates, cells were seeded at a concentration of 10,000 cells per well into the interior wells of the plates (columns 3-11, rows B-G inclusive) in 200 μL of serum-supplemented DMEM (MDA-MB-231 and MDA-MB-436 cells) or supplemented RPMI (A549 cells). Outer wells were filled with 200 μL of supplemented DMEM (MDA-MB-231 and MDA-MB-436 cells) or supplemented RPMI (A549 cells). Cells were incubated at 37 °C and 5% $CO_2$ for 24 hours before photosensitiser administration. After 24 hours of incubation, supplemented media (DMEM or RPMI) was replaced with photosensitiser-containing, serum-free DMEM at various concentrations and the plates incubated for 1 hour (Verteporfin), 3 hours (5-ALA), or 24 hours (Temoporfin) under low light conditions at 37 °C and 5% $CO_2$. The photosensitiser solution was then removed and cells were washed once with DPBS and serum-free DMEM was added to wells. Cells were then illuminated from below using LED arrays set to a power output of ~ 5 mW per LED for 15 minutes (Verteporfin) or 25 minutes (5-ALA and Temoporfin) (power output of 660 nm LEDs was limited to ~3.5 mW). 96-well plates were then incubated for 24 hours at 37 °C and 5% $CO_2$ under low-light conditions. 24 hours post illumination, 10 μL of cell-counting kit 8 (CCK-8) substrate (Sigma Aldrich) was added to each well and the plates incubated for 3 hours at 37 °C and 5% $CO_2$. After 3 hours, the absorbance of each well was measured at 450 nm using a plate reader (SpectraMax M5).

## LIVE/DEAD$^{TM}$ Assay

1 x 10$^6$ A549 cells were seeded into each of the wells of a 6-well plate in serum-supplemented RPMI and incubated for 24 hours at 37 °C and 5% $CO_2$. Cells were then incubated with serum-free DMEM with or without 10 mM 5-ALA for 3 hours at 37 °C and 5% $CO_2$. Solutions were then removed, and cells washed with DPBS. Serum-free DMEM was then added to all cells. Cells in each well were illuminated using the multimodal optical probe and associated lasers for Raman spectroscopy and PDT. The probe was placed underneath the centre of each well and held in place by a support stand. Laser outputs were adjusted to 100 mW and cells illuminated for 120 seconds with one of the lasers (no laser illumination for control cells) before the plate was returned to the incubator for 24 hours at 37 °C and 5% $CO_2$. After 24 hours incubation, a LIVE/DEAD$^{TM}$ assay was performed using a 2 μM calcein AM (Thermo Fisher Scientific) and a 4 μM ethidium homodimer-1 (Thermo Fisher Scientific) solution in DPBS. Media was removed from the cells and cells were washed with DPBS. LIVE/DEAD$^{TM}$ reagent was added to the cells and incubated for 30 minutes at room temperature. Reagent was removed from cells and cells washed with DPBS. Cells were then placed in DPBS for immediate imaging. Imaging was performed using an Olympus IX71 inverted fluorescence microscope with a 4x objective. Multiple images were manually stitched together for Figure 3a.

## Ethics Statement

All animal studies were approved by the University College London Biological Services Ethical Review Committee and licensed under the UK Home Office regulations and the Guidance for the Operation of Animals (Scientific Procedures) Act 1986 (Home Office, London, United Kingdom) and United Kingdom Coordinating Committee on Cancer Research Guidelines for the Welfare and Use of Animals in Cancer Research[74]. All in vivo experiments were performed under isoflurane anaesthesia (1.5% - 2.5% isoflurane in oxygen 1.5 L/min).

## In Vivo Raman-PDT Theranostics

Ten female CD1 *nu/nu* mice (6-8 weeks old, 25-30g) were subcutaneously injected with 1x10$^6$ cells from a human colorectal carcinoma cell line, SW1222, on their right flank. These cells were previously engineered in-house to express the luciferase gene for bioluminescence imaging (BLI)[75]. Mice were housed for 8 days to allow tumours to develop. Prior to 5-ALA administration, each tumour was imaged using BLI and 20 Raman spectra were collected from the tumour and the control flank for each mouse using the theranostic system described using the 785 nm laser with a 100 mW power output and 1 second integration time. Five mice received a 50 mg/kg tail vein injection of a 15 mg/mL 5-ALA solution while five control mice received no injection. At 1, 2, and 4 hours post 5-ALA administration, tumours were imaged using BLI and Raman spectra were obtained from the tumour and the control flanks of all mice as described. After the final Raman spectral acquisitions, mice were treated with the 633 nm PDT laser with a fluence rate of 50 mW/cm$^2$ and a total fluence of 30 J/cm$^2$. Mice were rehoused and monitored for 6 days post PDT. At 1, 3, and 6 days post PDT, tumours of control and PDT mice were reimaged using BLI, and Raman spectra were obtained from the tumour and the control flanks of all mice as described. At 6 days post PDT, the 5 PDT mice were re-injected with 50 mg/kg of a 15 mg/mL 5-ALA solution to enable quantification of PPIX tumoural uptake, but no additional PDT was performed. All mice were subsequently sacrificed by cervical dislocation, in accordance with local regulations, and tumours excised for PPIX quantification.

**Bioluminescence Imaging**

BLI was performed using IVIS Lumina (PerkinElmer, USA). Animals were administered 200 µL D-luciferin (Promega) intraperitoneally at 75 mg/kg. 2 or 3 mice were imaged simultaneously per acquisition. Mice were anaesthetised and sequential BLI images were acquired 5 minutes after luciferin injection using auto exposure time with 0.5 minutes delay between two consecutive acquisitions. A circular region of interest (ROI) was placed over the tumour on the first image and subsequently pasted over every new image acquired until all ROIs reach their maximum intensity. The total signal in the ROI was quantified as total flux (photons/s) using Living Image software version 4.5 (PerkinElmer). Representative images were presented using radiance (p/sec/cm$^2$/sr) as the colour scale utilising the same software.

**PPIX Tissue Quantification Assay**

SW1222 colorectal tumours for PPIX quantification were stored on ice under dark conditions prior to processing. Tumours were weighed before immersion in 1 mL of Solvable$^{TM}$ (PerkinElmer) and vials placed into an ultrasound bath at 45 °C until the tissue was dissolved (~ 4 hours). Three 50 µL aliquots were taken from each tumour solution and each diluted in 1mL of Solvable$^{TM}$ before being placed into the ultrasound bath for a further 15 minutes. For each tumour solution aliquot, 700 µL was placed into a plastic cuvette and the absorption spectrum between 350-750 nm recorded following excitation at 400 nm using a Horiba FL-1000 spectrofluorometer. Fluorescence spectra were cropped to between 450-750 nm, the background fluorescence spectrum of pure Solvable was subtracted, and then the tissue autofluorescence background was subtracted using a Whittaker filter ($\lambda = 10$) before filtering with a Savitzky-Golay filter (1$^{st}$ order, frame length = 7). A standard curve was developed using control tumour tissue, processed as above, with increasing known amount of PPIX in a Solvable solution added to the cuvette. Standard curve fluorescence spectra were processed as described above.

## Acknowledgements

C.C.H. acknowledges funding from the NanoMed Marie Skłodowska-Curie ITN from the H2020 programme under grant number 676137. M.S.B. acknowledges support from H2020 through the Individual Marie Skłodowska-Curie Fellowship "IMAGINE" (701713). A.N. and M.M.S. acknowledge support from the GlaxoSmithKline Engineered Medicines Laboratory. M.Z.T. and T.L.K. acknowledge funding from the EPSRC for the award of an Early Career Fellowship EP/L006472/1. I.J.P. acknowledges support from the Whitaker International Program, Institute of International Education, United States of America. U.K. acknowledges support from the Deutsche Forschungsgemeinschaft [KA 4370/1-1]. D.J.S. acknowledges support from a British Heart Foundation Intermediate Basic Science Research Fellowship (FS/15/33/31608), the MRC MR/R026416/1 and the Wellcome Trust. M.M.S. acknowledges a Wellcome Trust Senior Investigator Award (098411/Z/12/Z). We also acknowledge use of microscopy facilities within the Facility for Imaging by Light Microscopy (FILM) at Imperial College London, which is part-supported by funding from the Wellcome Trust (grant 104931/Z/14/Z). Raw data is available upon reasonable request from rdm-enquiries@imperial.ac.uk.

# Supplementary Information

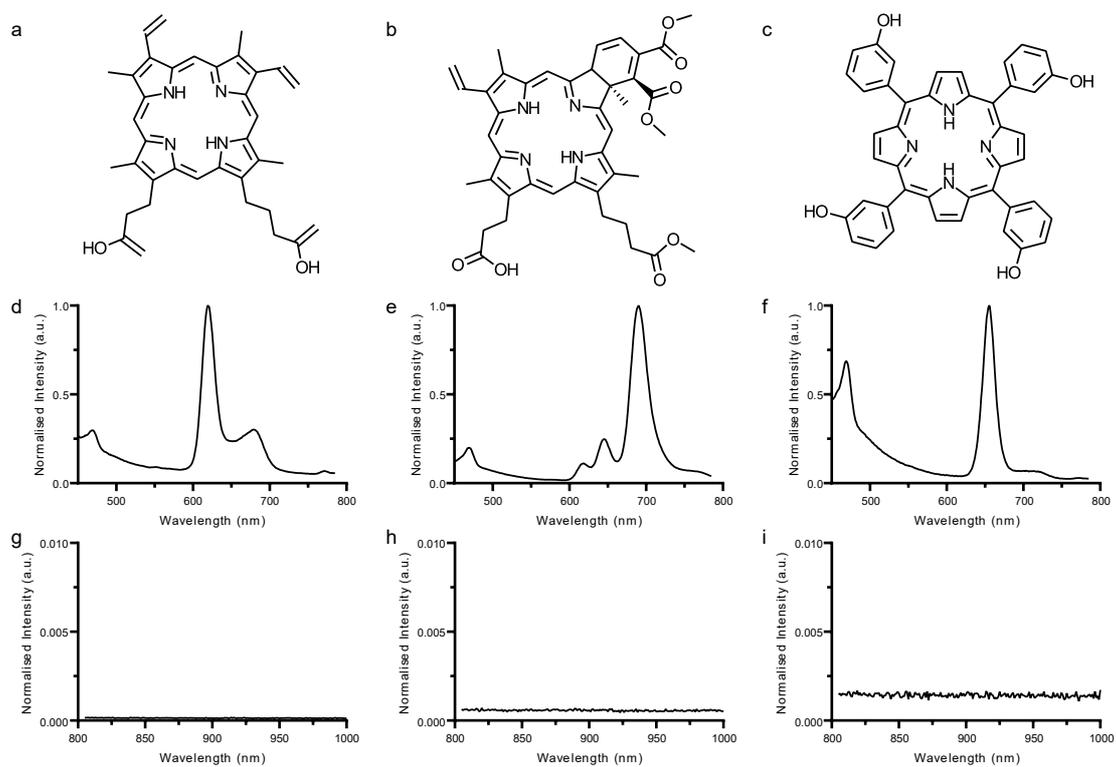

**Supplementary Figure 1 | Photosensitisers for Raman-PDT Theranostics.** (**a-c**) Chemical structures of clinically employed photosensitisers investigate for Raman-PDT theranostic system; (**a**) Protoporphyrin IX (PPIX), (**b**) Verteporfin, (**c**) Temoporfin. (**d-f**) Normalised fluorescence emission spectra (ex 405 nm) of (**d**) PPIX, (**e**) Verteporfin, (**f**) Temoporfin. (**g-i**) Normalised fluorescence emission spectra (ex 785 nm) of (**g**) PPIX, (**h**) Verteporfin, (**i**) Temoporfin.

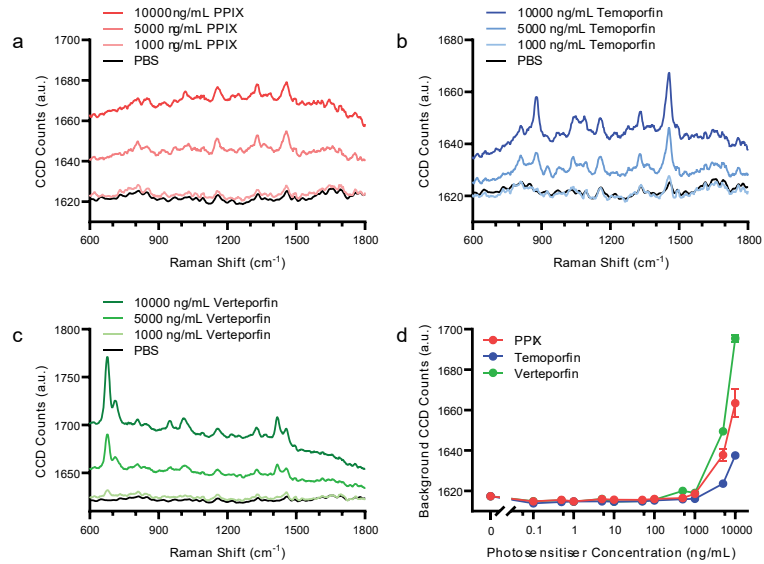

**Supplementary Figure 2 | Raw Raman Spectra of Photosensitiser Solutions.** (**a-c**) Raw Raman spectra of (**a**) PPIX, (**b**) Temoporfin, and (**c**) Verteporfin serial dilutions as compared to PBS (n = 5). Major peaks seen in (**b**) and (**c**) correspond to background traces of solvents used in preparation of Temoporfin and Verteporfin solutions. (**d**) Peak fluorescence backgrounds for photosensitizer serial dilutions (mean ± S.D., n = 5).

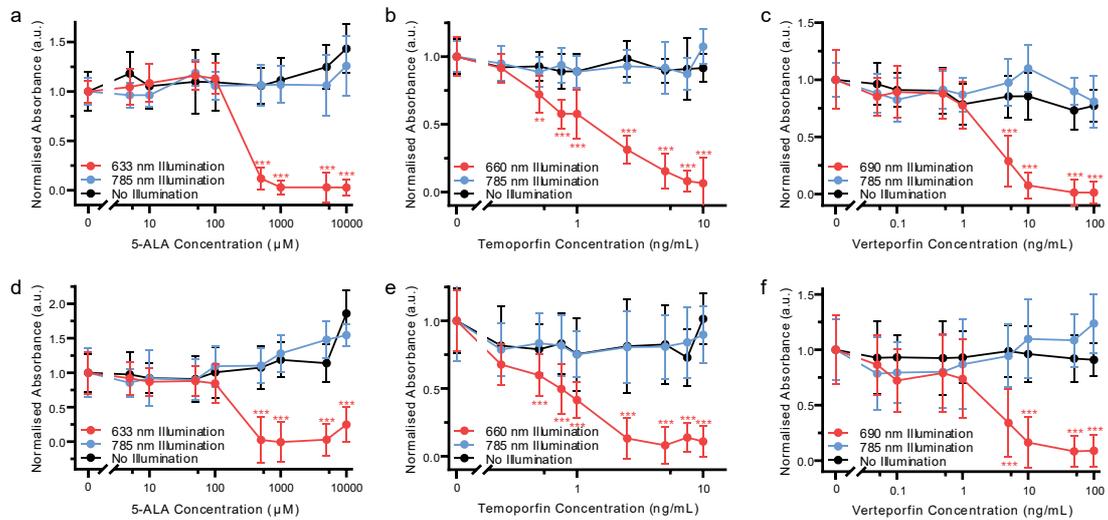

**Supplementary Figure 3 | Photosensitiser Cell Viability Assays.** (**a-c**) Cell viability assays of MDA-MB-231 cells incubated with (**a**) 5-ALA, (**b**) Temoporfin, (**c**) Verteporfin. (**d-f**) Cell viability assays of MDA-MB-436 cells incubated with (**d**) 5-ALA, (**e**) Temoporfin, (**f**) Verteporfin. (mean ± S.D., N = 3, n = 6) (Error bars: mean ± STD) (Multiple comparisons *t*-test, Bonferroni post hoc correction, * $P < 0.05$, ** $P < 0.01$, *** $P < 0.001$).

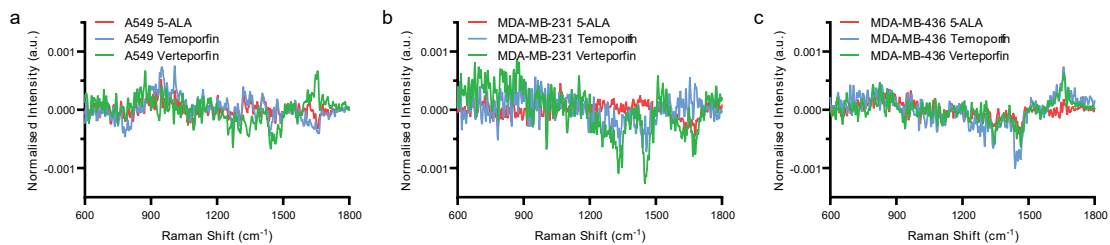

**Supplementary Figure 4 | Raman Difference Spectra of Photosensitiser Cells.** (**a-c**) Raman difference spectra (10 s integration time) of cells in the presence of different photosensitisers (phenol red-free DMEM (Control), 5-ALA (10000 µM), Verteporfin (100 ng/mL), or Temoporfin (10 ng/mL)), calculated as 'PS Cell – Control Cell' for (**a**) A549 cells, (**b**) MDA-MB-231 cells, and (**c**) MDA-MB-436 cells (N = 10, n = 5).

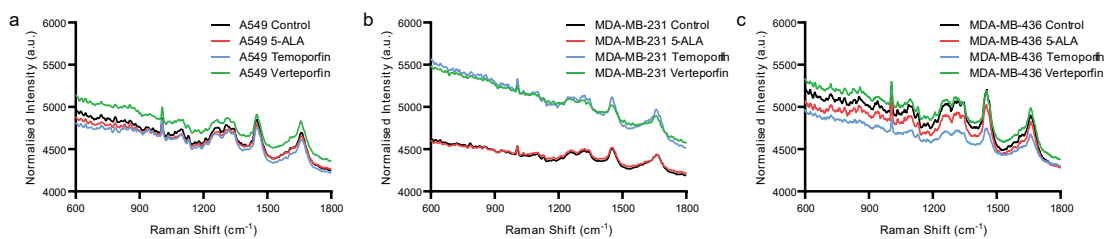

**Supplementary Figure 5 | Raw Raman Spectra of Photosensitiser Cells.** (**a-c**) Raw Raman spectral acquisitions (10 s integration time) of (**a**) A549 cells, (**b**) MDA-MB-231 cells, and (**c**) MDA-MB-436 cells in the presence of different photosensitisers (phenol red-free DMEM (Control), 5-ALA (10000 µM), Verteporfin (100 ng/mL), or Temoporfin (10 ng/mL)) (N = 10, n = 5).

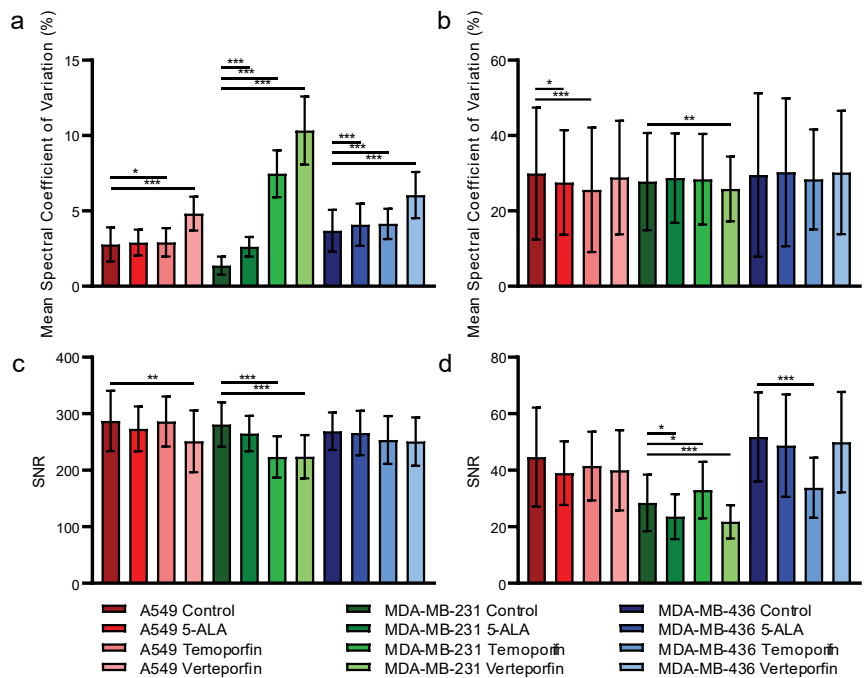

**Supplementary Figure 6 | Mean Spectral Coefficient of Variation and Signal-to-Noise Ratio of Photosensitiser Cells.** (a-b) Mean spectral coefficient of variation of (a) raw and (b) processed Raman photosensitiser cell spectra. (c-d) Mean SNR of (c) raw and (d) processed Raman photosensitiser cell spectra (N = 10, n = 5) (Error bars: mean ± STD) (Two-way analysis of variance (ANOVA), Tukey's honest significant differences (HSD) post hoc correction, * $P < 0.05$, ** $P < 0.01$, *** $P < 0.001$).

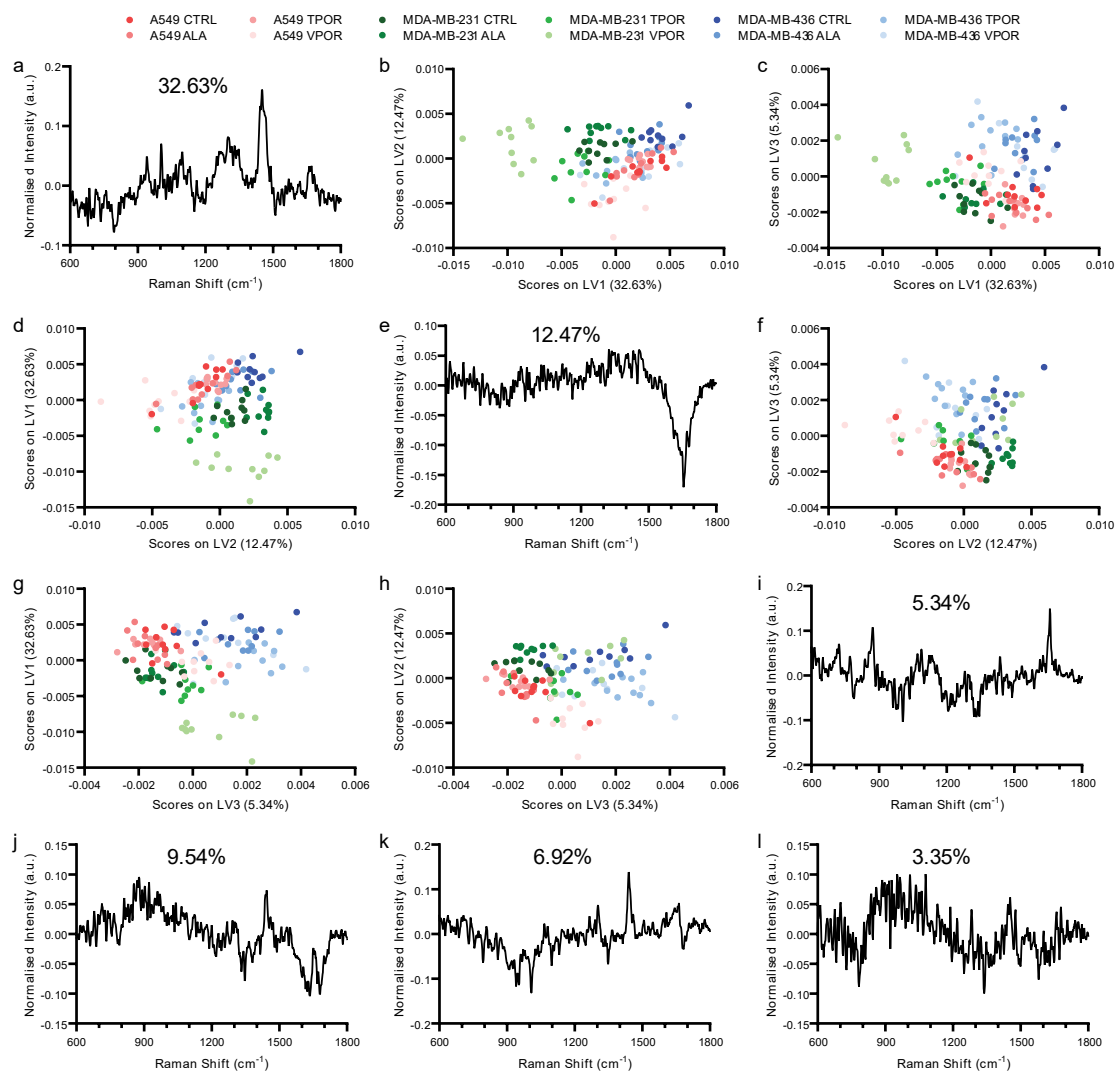

**Supplementary Figure 7 | Photosensitiser Cell Raman Spectra PLS-DA.** (**a-i**) Matrix plot of (**a, e, i**) latent variables 1-3 for PLS-DA of processed Raman spectra performed across the three cell lines, A549, MDA-MB-231, and MDA-MB-436 (blind to the presence or absence of different photosensitisers) (N = 40, n = 5). (**j-l**) PLS-DA latent variables 4-6. Percentages indicate percentage variance explained by each latent variable.

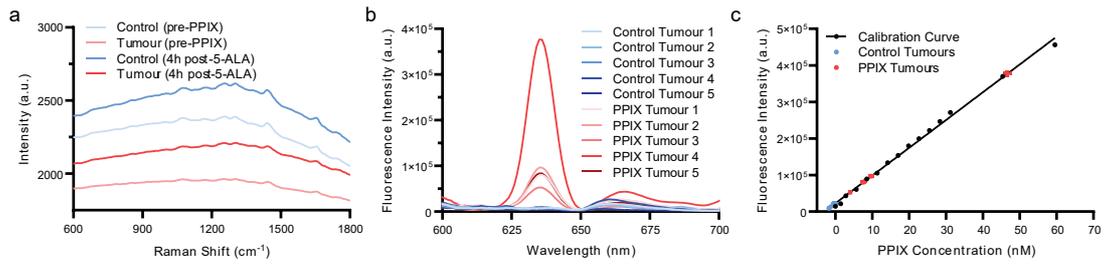

**Supplementary Figure 8 | Confirmation of PPIX Uptake SW1222 Tumours *In Vivo*.** (**a**) Mean raw Raman spectra of control flanks and tumours in mice pre-5-ALA induced PPIX and 4 hours post-5-ALA injection (50 mg/kg) (n = 18-20, N = 5). (**b**) Emission spectra of control tumours and PPIX positive tumours following re-administration of 5-ALA (50 mg/kg) with a 4-hour incubation time 6 days post PDT treatment immediately prior to tumour excision. (**c**) Quantification of PPIX tumour concentration for control and PPIX positive tumours.

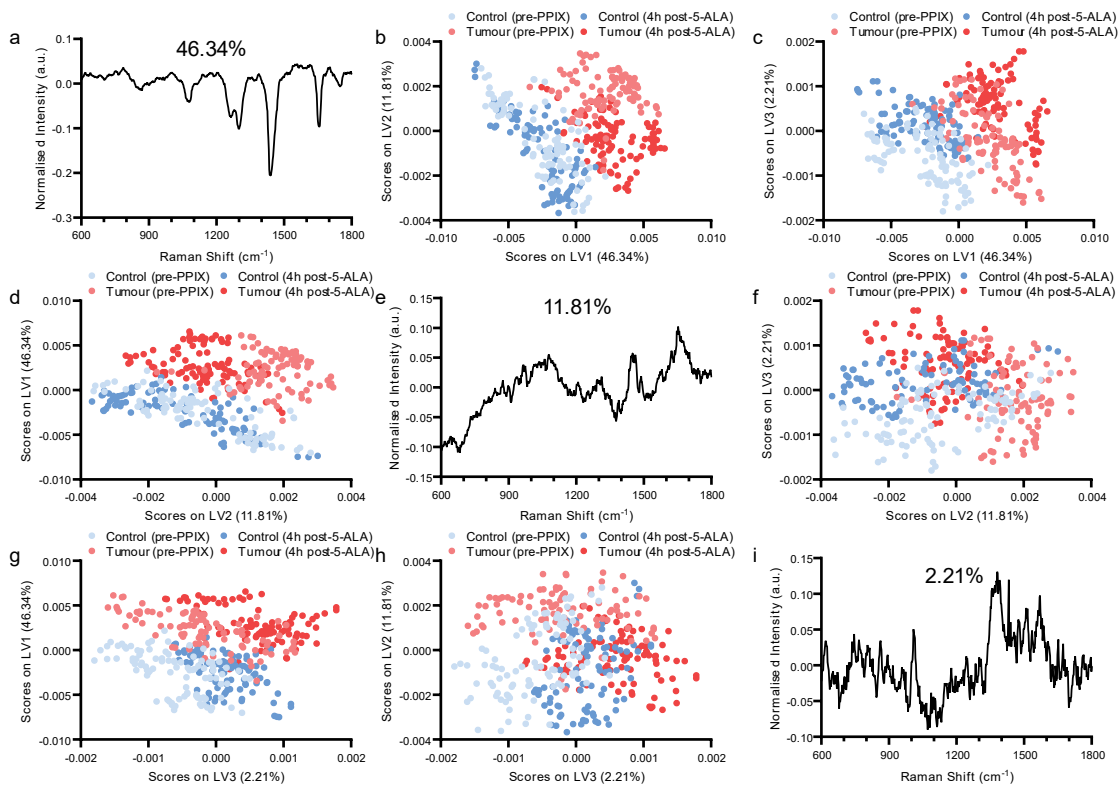

**Supplementary Figure 9 | PPIX+ SW1222 Tumours Raman Spectra PLS-DA.** (**a-i**) Matrix plot of (**a, e, i**) latent variables 1-3 for PLS-DA of processed Raman spectra for control tissue and tumour tissue pre-5-ALA induced PPIX and 4h post 5-ALA injection (50 mg/kg) (n = 18-20, N = 5). Percentages indicate percentage variance explained by each latent variable.